\documentclass[journal=jctcce,manuscript=article]{achemso}
\setkeys{acs}{articletitle=true}

\usepackage[version=3]{mhchem} 
\usepackage[T1]{fontenc}       
\usepackage{graphicx}
\usepackage{gensymb}
\usepackage{caption}
\usepackage{subcaption}
\usepackage{wrapfig}
\usepackage{multirow}
\usepackage{hhline}

\author{Daryna Smyrnova}
\affiliation[Quantum Chemistry and Physical Chemistry Division, Department of Chemistry , KU Leuven, Celestijnenlaan 200F, 3001 Heverlee, Belgium]
{Quantum Chemistry and Physical Chemistry Division, Department of Chemistry , KU Leuven, Celestijnenlaan 200F, 3001 Heverlee, Belgium}
\author{Mar\'{i}a del Carmen Mar\'{i}n}
\affiliation[Department	of	Biotechnology,	Chemistry	e	Pharmacy,	Universit\'{a}	di	Siena,	via	A.	Moro	2,	I-53100	Siena,	Italy] 
{Department	of	Biotechnology,	Chemistry	e	Pharmacy,	Universit\'{a}	di	Siena,	via	A.	Moro	2,	I-53100	Siena,	Italy}
\alsoaffiliation[Department	of	Chemistry,	Bowling	Green	State	University,	Bowling	Green,	OH	43403,	USA]
{Department	of	Chemistry,	Bowling	Green	State	University,	Bowling	Green,	OH	43403,	USA}
\author{Massimo Olivucci}
\affiliation[Department	of	Biotechnology,	Chemistry	e	Pharmacy,	Universit\'{a}	di	Siena,	via	A.	Moro	2,	I-53100	Siena,	Italy]
{Department	of	Biotechnology,	Chemistry	e	Pharmacy,	Universit\'{a}	di	Siena,	via	A.	Moro	2,	I-53100	Siena,	Italy}
\alsoaffiliation[Department	of	Chemistry,	Bowling	Green	State	University,	Bowling	Green,	OH	43403,	USA]
{Department	of	Chemistry,	Bowling	Green	State	University,	Bowling	Green,	OH	43403,	USA}
\author{Arnout Ceulemans}
\affiliation[Quantum Chemistry and Physical Chemistry Division, Department of Chemistry , KU Leuven, Celestijnenlaan 200F, 3001 Heverlee, Belgium]
{Quantum Chemistry and Physical Chemistry Division, Department of Chemistry , KU Leuven, Celestijnenlaan 200F, 3001 Heverlee, Belgium}
\email{arnout.ceulemans@kuleuven.be, daryna.smyrnova@kuleuven.be}

\title[trend in RSFPs fluorescence]
  {On excited state reaction path in reversibly switchable fluorescent proteins}

\abbreviations{RSFP, PMF, MFEP, FEP, US}


\begin{document}
\begin{tocentry}
\begin{center}
\includegraphics[width=9cm]{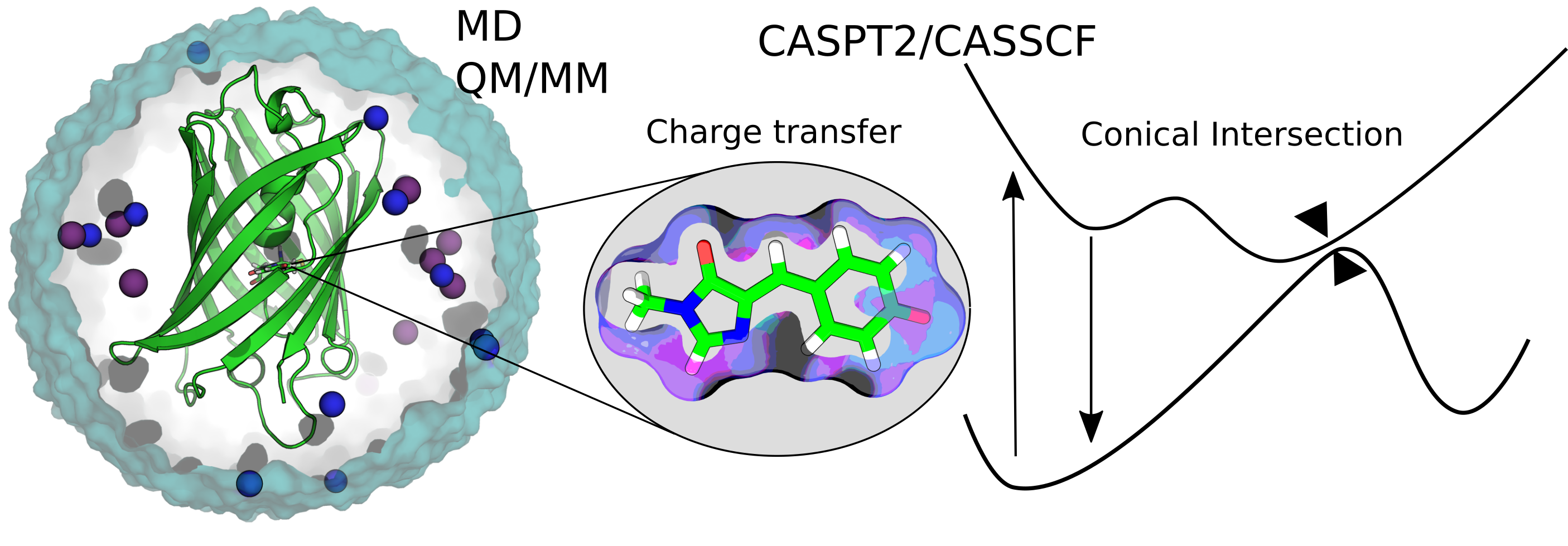}
  \label{fgr:For Table of Contents Only}
\end{center}
\end{tocentry}

\newpage
\begin{abstract}
A set formed by five reversibly-switchable fluorescent proteins (RSFPs) display spread over 40~nm in absorption maxima and only 18~nm in emission. The five proteins -- Dronpa, rsFastLime, rsKame, Padron(anionic form) and bsDronpa -- carry exactly the same chromophore and differ just in a few mutations. Thus they form an ideal set for mechanistic investigation. Starting with the results of molecular dynamics simulations we use QM/MM calculations to investigate the effects controlling the spectral tuning . In this contribution we show that the models, which are based on CASPT2//CASSCF level of QM theory, reproduce the observed absorption trend with only a limited blue-shift of 4.5 kcal/mol and emission trend with even smaller blue-shift of 1.5 kcal/mol. Using CASSCF QM/MM calculations we analyze the chromophore's charge-transfer patterns during the absorption and emission, which, in turn trigger a cascade of hydrogen-bond-network rearrangements indicating  a preparation to an isomerization event.  We also show how the contribution of individual aminoacids to the chromophore conformational changes correlates with spectral tuning of the absorption and emission. Furthermore, we identify how the conical intersection topography correlates with protein's photophysical properties. In conclusion, we establish a detailed mechanistic explanation of variations in photo-switching speed as well as higher sensitivity of RSFPs to mutation observed for light absorption relative to light emission.
\end{abstract}

\section{Introduction}

The broad range of applications for Green Fluorescent Proteins (GFPs)\cite{ grynkiewicz_new_1985,shaner_improved_2004-1,shimomura_extraction_1962} keeps them in the spotlight of researchers for many years now. However, it seems that once you think you have achieved almost perfect fluorescent proteins\cite{okabe_green_1997,lelimousin_intrinsic_2009}, new applications arise spanning a completely new field in fluorescent proteins research\cite{fernandez-suarez_fluorescent_2008, dedecker_fluorescent_2013,subach_chromophore_2012,kent_deconstructing_2008}. One such breakthrough was the discovery of the Reversibly-Switchable Fluorescent Proteins (RSFPs)\cite{habuchi_photo-induced_2006,habuchi_cover:_2005,hofmann_breaking_2005}. They are an indispensable tool in super-resolution microscopy\cite{dedecker_subdiffraction_2007, vandenberg_diffraction-unlimited_2015} and improvements to their photophysical properties are still being carried on \cite{bourgeois_reversible_2012,acharya_photoinduced_2016}. Theoretical studies have contributed significantly to understanding of RSFPs fluorescence quantum yield\cite{lelimousin_intrinsic_2009}, modulated isomerization quantum yield\cite{groenhof_photoactivation_2004} and color-tuning\cite{shcherbakova_red_2012}.

In spite of the large amount of research in this field, it seems like we still do not have a complete answer to the question why some proteins are photo-switchable while others are just fluorescent. How can only a few mutations in the protein $\beta$-barrel shift the absorption maxima by 40~nm\cite{andresen_photoswitchable_2008, faraji_nature_2015}? What is the role of the protein environment in this process and most importantly how does each aminoacid contribute to the photophysical properties of a certain protein?

There are a number of studies conducted on various fluorescent proteins identifying the role of protein environment on absorption maxima\cite{amat_spectral_2013} as well as emission and various electrostatic effects\cite{hasegawa_excited_2007, park_emission_2016}. Study on a set of green, orange and red FPs by Hasegawat et al.\cite{hasegawa_excited_2007} identified the blue-shifting effect of the protein electrostatic potential. Another detailed study on DsRed by List et al.\cite{list_molecular-level_2012} identifies the role of protein environment in one- and two-photon absorption. Also theoretical models were built to explain the pressure and temperature effects \cite{jacchetti_temperature_2016} or electrostatic field spectral tuning in GFPs\cite{drobizhev_long-_2015}. Studies on fluorescence and potential energy landscape of anionic GFP chromophore by Martin et al\cite{martin_origin_2004}  and Polyakov et al\cite{polyakov_potential_2010} provide in-depth mechanistic details for possible chromophore isomerization pathways. But in order to have an outlook which would allow for a rational protein design, a study on a consistent set of proteins is indispensable. That is why we have chosen a set of proteins differing by a minimal number of mutations, while conserving the chromophore structure. Step-by-step exploration of an excited state surface allowed us to connect absorption/emission, photo-switching speed and conical intersections with protein environment effects.

A well-known RSFP Dronpa\cite{habuchi_photo-induced_2006,moeyaert_green--red_2014} which was a pioneer in RSFPs field and is still widely used for various applications\cite{dedecker_fluorescent_2013,zhou_photoswitchable_2013}, became constitutive of the model protein set. Also, due to its pioneering role many Dronpa-like mutants were developed and well characterized\cite{andresen_photoswitchable_2008}. For our study we have selected Dronpa mutants which do not alter the chromophore structure, so all of the mutations occur in the $\beta$-sheets and/or $\alpha$-helix. The list of five proteins that were used in our study in shown in the Table \ref{tab:mutations}. 

\begin{minipage}{\linewidth}
\centering
  \begin{tabular}{lllll}
    \hline
    Protein&Mutation&Absorption&Emission\\
    &&nm/kcal*mol\textsuperscript{-1}&nm/kcal*mol\textsuperscript{-1}\\
    \hline
    Dronpa\cite{moeyaert_green--red_2014}&&503/56.7&522/54.7\\ 
    rsFastLime\cite{stiel_1.8_2007}&V157G&496/57.5&518/55.1\\
    rsKame\cite{rosenbloom_optimized_2014}&V157L&503/56.7&515/55.4\\
    Padron0.9\cite{andresen_structural_2007, brakemann_molecular_2010}&T59M+V60A+N94I+P141L&504/56.6&510/56.0\\
    &G155S+V157G+M159Y+F190S&&\\
    bsDronpa\cite{andresen_photoswitchable_2008}&A69T+D112N+G155S+V157G&460/62.0&504/56.6\\
    &M159C+F173C&&\\
    \hline
  \end{tabular}\par
\captionof{table}{Effect of mutations on the absorption and emission wavelength in Dronpa-like mutants. Only the most essential mutations are indicated for Padron0.9. For bsDronpa six essential mutations are shown, which also were used to model it on the basis of Dronpa crystal structure.}
 \label{tab:mutations} 
  \bigskip
\end{minipage}

Two single-point mutants rsKame\cite{rosenbloom_optimized_2014} and rsFastLime\cite{stiel_1.8_2007, andresen_photoswitchable_2008} were chosen due to their minimal difference with Dronpa. They do have a big difference in the speed of photo-switching (see Table S1), but there is barely any difference in the absorption and emission maxima. For a broader spectral variation we have chosen Padron0.9\cite{brakemann_molecular_2010} and bsDronpa\cite{andresen_photoswitchable_2008}. Both proteins conserve the chromophore structure with only a few mutations in the protein matrix. Padron0.9 is a negatively-switchable counterpart of Dronpa. But more importantly for us its absorption and emission maxima are slightly red-shifted in comparison to Dronpa. And bsDronpa is the most blue-shifted close mutant of Dronpa with a large Stokes shift (43~nm). Previous studies on FPs with large Stokes shift\cite{faraji_nature_2015, piatkevich_extended_2013, yoon_far-red_2016} have shown that the extended Stokes shift is a result of a more flexible hydrogen bond (H-bond) interplay where more states can be accessed in the ground and/or excited state surface. However, here our focus is on obtaining one bsDronpa population capable of reproducing the spectral trend of our model set. Hence we do not indulge into in depth conformational analysis of bsDronpa H-bonds fluctuations. 

We would like to emphasize that the proteins used in our calculations differ by a maximum of 8 mutations and stem from the same protein thus minimizing the differences in the environment. Also, they contain exactly the same chromophore including its sidechain. Although the chromophore's $\pi$-system includes only phenol and imidazoline rings, the chromophore sidechain also plays an important role influencing the $\alpha$-helix flexibility and adaptability of the chromophore to the protein environment\cite{arpino_crystal_2012, moeyaert_green--red_2014, smyrnova_thermal_2016}.  Since we are interested in extrapolation of our results into functional mutations, conserving the protein structure is of a an extreme importance. The more mutations are included in the protein the less predictable become the results with increasing possibility of a failed experiment due to factors which are out of reach for computational studies.

Previous studies have shown how choice of an initial set up of the system can influence the calculated excitation energies\cite{filippi_bathochromic_2012,melaccio_towards_2016,amat_spectral_2013}. Here we have used exactly the same protocol for all five proteins, which can be briefly described as a modified Automatic Rhodopsin Modelling(ARM) protocol developed by Melaccio et al\cite{melaccio_towards_2016}. Starting QM/MM structures were taken from an extensive MD simulations. Then after CASSCF QM/MM optimization  CASPT2  method was used to calculated excitation and emission energies for a set of ten structures. Representative structures for each protein were chosen for further reaction pathway and charge transfer analysis. Charge transfer in the chromophore during excitation and emission is analyzed in correlation with the conformational changes in protein environment. Also, through calculation of individual aminoacid contributions to absorption/emission, we provide an explanation for a remarkable blue-shift in bsDronpa absorption spectra. And finally conical intersection and a closest lying minimum was identified for each of the proteins. As a result an excited state reaction pathway is assessed in a rational and unifying manner. 

\newpage
\section{Results}
\subsection{Reproducing the experimental absorption and emission values}

\begin{figure}[H]
\includegraphics[width=13.5cm]{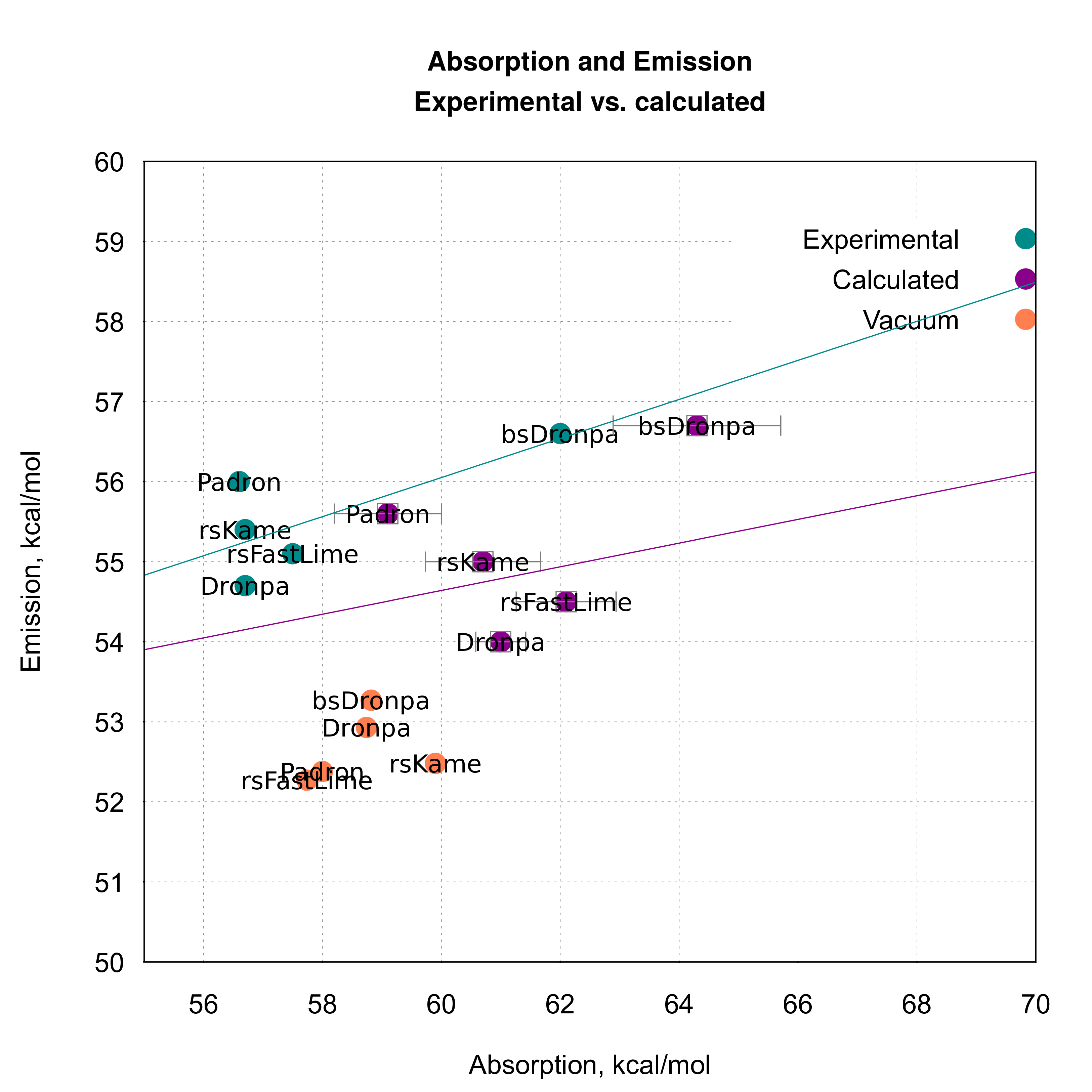}
  \caption{Comparison of the calculated (violet) and experimental (cyan) absorption and emission values for the model-set of five proteins. Results for the respective chromophores in vacuum are shown in orange. For the calculated values error bars are shown as well.}
  \label{fgr:trend}
\end{figure}

Results for the average absorption and emission values obtained in our study are shown in the Figure \ref{fgr:trend}. First of all, we would like to note that the absorption and emission trend is completely reproduced in our results. Not only a shift absorption/emission values is rather small, but also the order of proteins is conserved. Standard deviation of the average emission values with respect to the experimental emission values is around 1.5 kcal/mol (for all proteins). As it can be seen from the Figure \ref{fgr:trend} calculated absorption values are more blue-shifted than emission values (origins of that phenomena are discussed further in our paper). It is reflected in the standard deviation of the absorption average from the experimental values - from 2.66 kcal/mol (Padron0.9) to 4.63 kcal/mol (bsDronpa).

It can be noticed that mutations introduced in Dronpa (see Table \ref{tab:mutations}) can significantly affect the absorption maxima, while emission maxima are much less spread. There are two main sources affecting the absorption and emission maxima: the chromophore conformation and the protein environment. Since the chromophore structure is conserved in all five proteins we see the major source for absorption distribution in the way protein environment interacts with the chromophore in the ground and excited state. To discard the influence of the chromophore conformation we have calculated the absorption and emission values for the chromophore in vacuum (orange points in the Figure \ref{fgr:trend}). Since the trend is drawn on the average value over ten structures for each protein, we have used the most representative (closest to the mean values) conformation. As expected neither trend nor significant spread in the values can be found in these results. Hence it can be concluded that we should focus on the protein environment effect on the absorption and emission.

\subsection{Charge re-distribution in the chromophore. Origins of the higher absorption sensitivity to the mutations}
One of the most intriguing features of our model set is a 6 kcal/mol absorption maxima difference between Padron0.9 -- bsDronpa and  only 2 kcal/mol difference in emission between Dronpa -- bsDronpa.  Not only absorption is much more sensitive to mutations, but also there is no correlation with the emission. In another words, a higher absorption wavelength does not imply a higher emission maximum.

\begin{figure}[H]
\includegraphics[width=15.5cm]{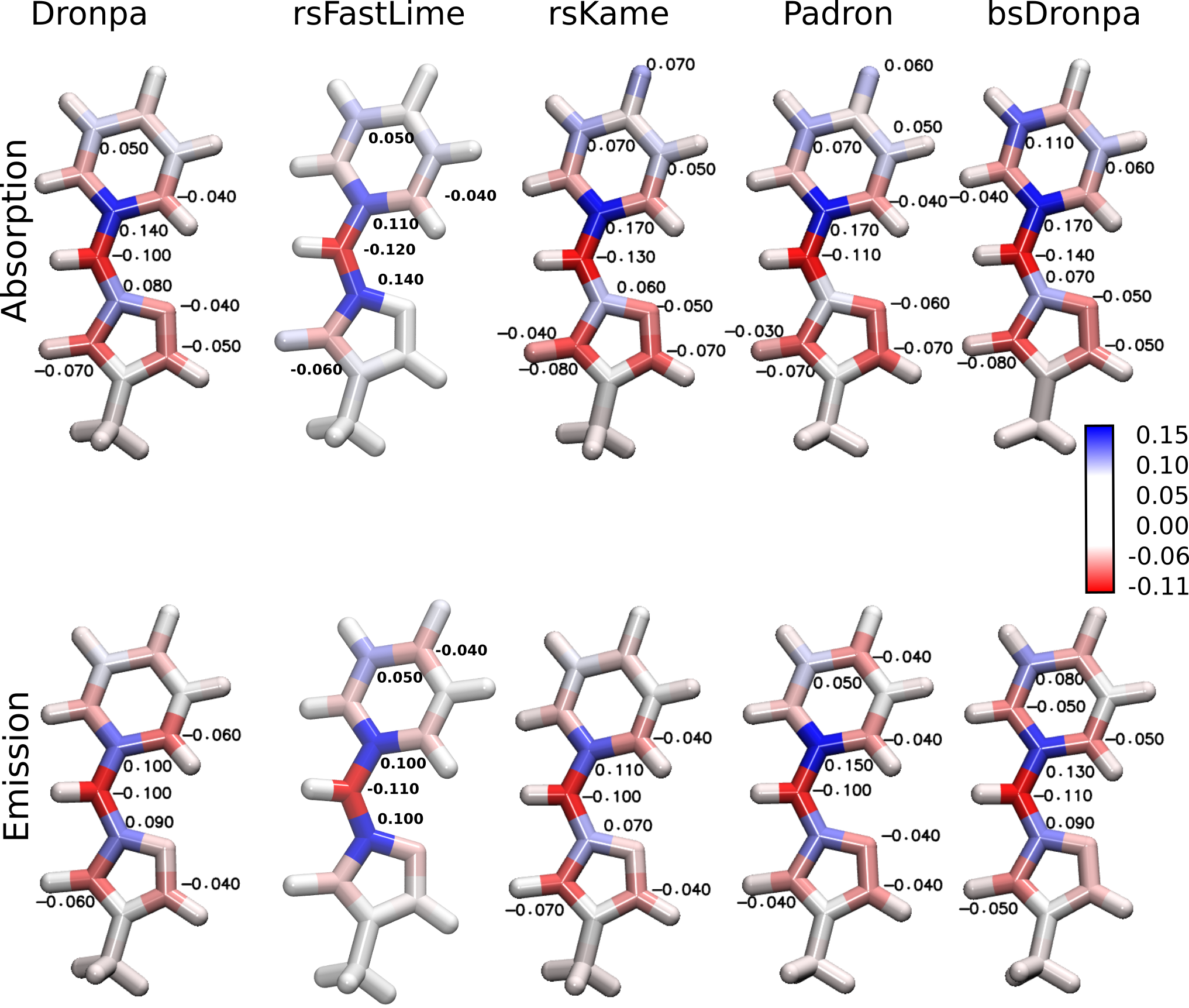}
  \caption{Comparison of the charge redistribution in the chromophore during absorption and emission. The images are color coded with the color bar shown on the right. Values are indicated next to the atoms for which the charge difference exceeds 0.04. Hydrogen charges are summed into the heavy atoms.}
  \label{fgr:charge_redistribution}
\end{figure}

We have hypothesized that the electrostatic effects may play a pivotal role in this phenomena. Hence we have looked into the charge redistribution during the excitation and emission. Then the same calculation was repeated for chromophore $\pi$-system, but already in vacuum. The values presented in the Figure \ref{fgr:charge_redistribution} correspond to the charge redistribution in protein environment during the absorption and emission. The charge redistribution is essentially an atomic charge difference between the first excited and ground state. 

Blue and red colors in the Figure \ref{fgr:charge_redistribution} correspond to a positive and a negative charge differences. It can be noticed that the negative charge is transferred from the the aromatic phenol ring to imidazoline ring and the bridging atoms of the single-double bond bridge. This observation is in line with Filippi et al.\cite{filippi_bathochromic_2012}  results on anionic GFP. In the ground state $\tau$-bond has a prevalent double character while $\phi$ is a single bond (benzenoid resonance structure). Observed charge redistribution corresponds to a double-bond delocalization over $\tau$ and $\phi$ thus facilitating the isomerization (or shift to a quinoid resonance structure). 

Although, charge redistribution pattern is similar during the absorption and emission the net charge transfer from phenol to imidazoline is more pronounced during the absorption (phenol ring is more blue and imidazoline ring is more red). When compared to the charge re-distribution in chromophore in vacuum no such difference was observed (see Figure~S1). Hence higher magnitude of the values spread in the absorption and not in the emission comes purely from protein environment.  First of all, mutations in the protein have direct effect on electrostatic potential around the chromophore. Also, pure spatial arrangement of the aminoacids around the chromophore can have important effects. In case when only one aliphatic aminoacid is mutated into another aliphatic one (Val 157 $\rightarrow$ Gly 157) minimal perturbation in electrostatic field around the chromophore is expected. However, slight rearrangement of the aminoacids around the chromophore can bring polar aminoacids closer to the chromophore and either red or blue-shift its absorption (individual influence of the aminoacids on spectral maxima is discussed further). 

Charge transfer also results in a slight conformational change. The geometrical parameters that change the most in the proteins are depicted in the Figure~S2. Change in all of the parameters indicates that the chromophore is preparing for an isomerization. The negative charge is transferred from the phenol ring into the imidazoline ring and particularly the bridging carbon atom. This leads to phenol ring -- H-bonds weakening. Also the H-bonds between Arg~89, Arg~64 and imidazoline ring get stronger. Thus leading to a slight Glu~144 -- His~193 H-bond elongation and a consequent Glu~144 -- His~193 bond shortening. Finally, the double bond bridge is elongated indicating transfer to a single bond character. As a consequence there is a noticeable out-of-plane twisting around $\tau$ dihedral angle. It has been shown earlier by Morozov et al.\cite{morozov_hydrogen_2016} that a lower number of H-bonds with the chromophore induces the isomerization. A link between this observation and a cascade of hydrogen-bond-network (HBN) transformations triggered after the excitation is shown in our study.


\subsection{Protein environment influence on the absorption and emission maxima}

In order to see influence of the individual aminoacids on the absorption and emission maxima, we have performed calculations where their charges were consecutively turned off.  Results of the calculation can be see in the Figure S3.

The importance of His~193 due to it's $\pi$-stacking interaction with the phenol ring of the chromophore. H193T mutation as reported by Li et al\cite{li_primary_2010} shows how absence of this interaction induces the internal  conversion and leads to fluorescence loss. Interaction of the chromophore with Tyr~116 is often overlooked and our calculations clearly demonstrate its importance. To our knowledge there is no experimental data published for Dronpa Y116 mutants, however Y116Q and Y116N mutations in pcDronpa (green-to-red photoconvertible mutant of Dronpa)\cite{moeyaert_green--red_2014} lead to an increased photo-conversion quantum yield and higher pKa. The blue-shifting effect of Arg~89 clearly manifests itself in bsDronpa, where it loses interaction with the chromophore. Although, for a definitive answer an X-ray structure of bsDronpa must be awaited for we anticipate that a large Stokes shift occurs due to Arg~89 flexibility. When its interaction with the chromophore is switched off both in theory (for Dronpa -- rsKame -- rsFastLime -- Padron) and in practice (bsDronpa) a blue-shift in absorption is observed. The importance of Arg~64 and Arg~89 can be attributed to their strong interaction with the chromophore through HBN. Both Arg~64 and Arg~89 for H-bonds with imidazoline ring of the chromophore. When absent, the negative charge transfer is impeded thus increasing the energy of the excited state. Consecutively a blue shift occurs.


\subsection{Conical intersections on the isomerization path}

Once we had a working protocol to study photophysical properties of the fluorescent proteins we decided to go further and connect their photochemical reactivity with the properties of conical intersections.

Conical intersections (CIs) very likely act as bottlenecks for the isomerization reactions in photo-activatable proteins. In particular, properties of the CI region should be crucial for photoswitching speed of the RSFPs. However in order to prove this an assessment of a group of photoswitching proteins is needed. At the same time there is an inverse correlation between the speed of photoswitching and fluorescence quantum yield (as seen in the Table S1). That suggests a pivotal role of excited state reaction kinetics due to possible presence of exited state energy barriers as well as distinctive topography of the conical intersection. 

To identify conical intersections on the excited state isomerization path of RSFPs we have used a conical intersection optimization with a constrain module available in Molcas\cite{fdez._galv?n_analytical_2016}. CI optimizations were carried out at CASSCF(12,11)/6-31G* level and afterwards a single-point SA-CASPT2(12,11)/6-31G* calculation was done on optimized structures. 

\begin{figure}[H]
\begin{subfigure}[b]{0.55\textwidth}
\includegraphics[width=9.1cm]{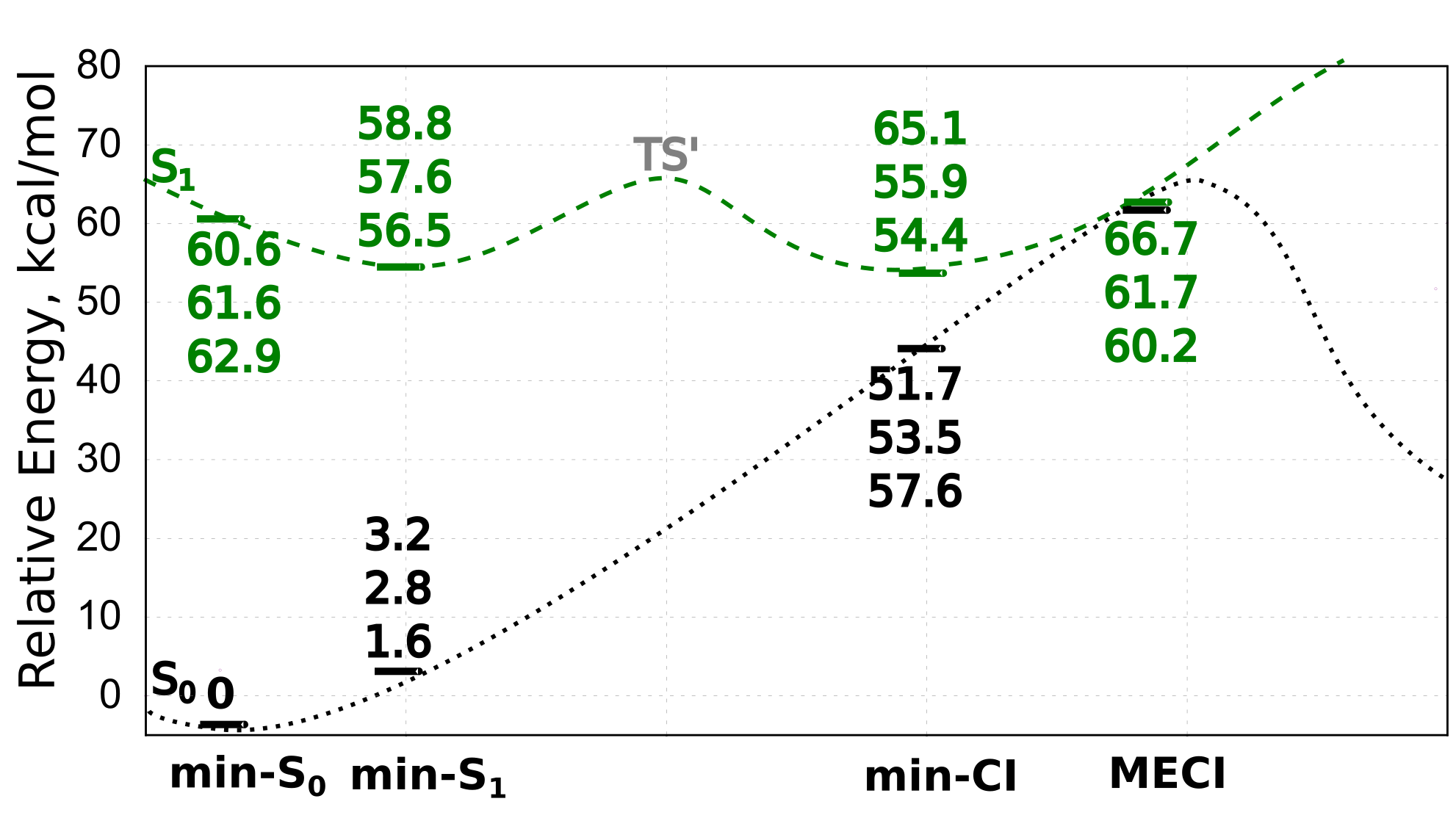}
\end{subfigure}
\begin{subfigure}[b]{0.40\textwidth}
\includegraphics[width=8.5cm]{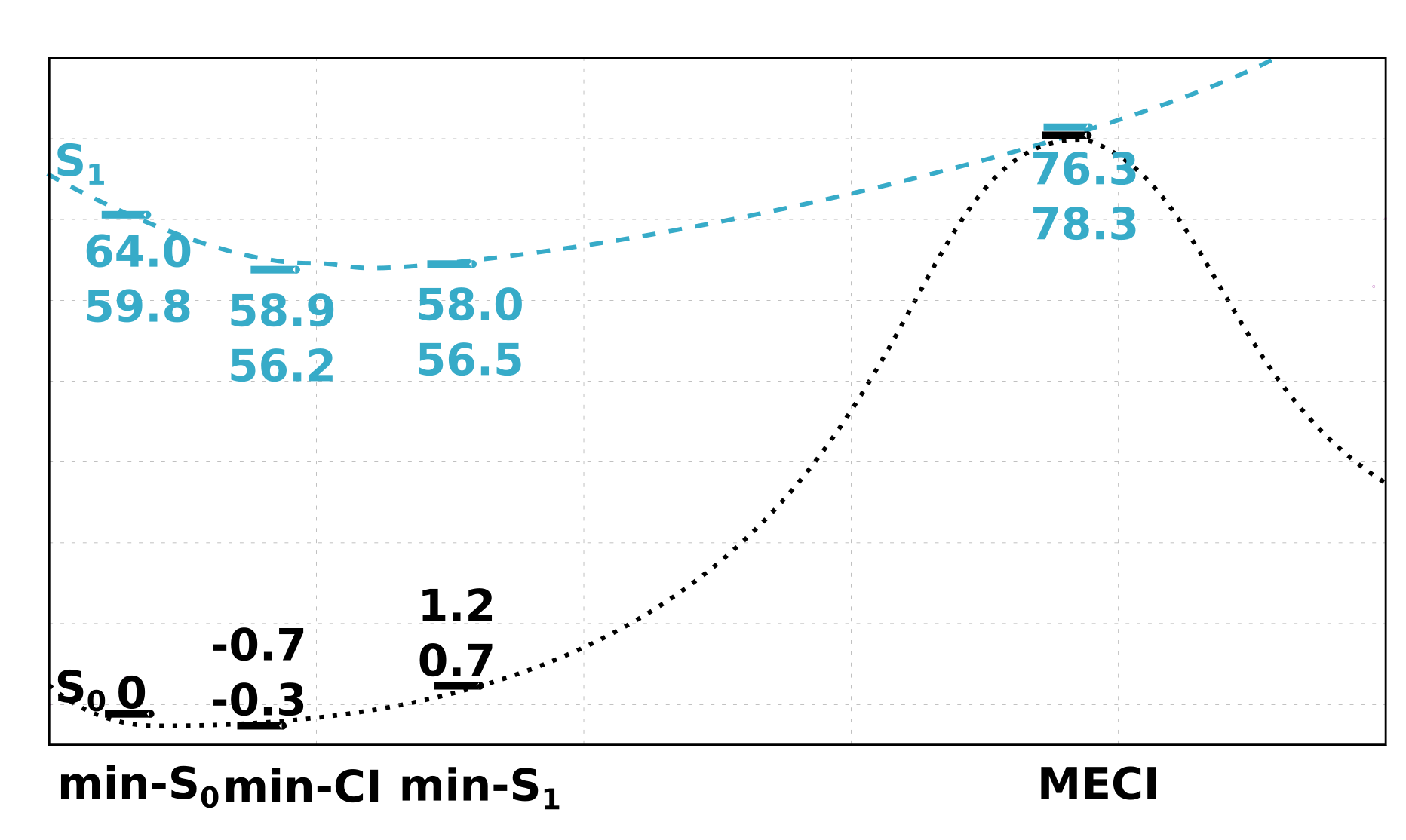}
\end{subfigure}
  \caption{Potential energy surface of the chromophore isomerization. Black lines correspond to the mechanism proposed for rsKame -- Dronpa --rsFastLime and blue lines are for bsDronpa -- Padron. The energy values are indicated in the corresponding order and color.  Dashed and continuous lines correspond to the ground and excited state surfaces respectively. TS region is indicated in grey in it's hypothetical position.}
  \label{fgr:CI}
\end{figure}

Results of our findings can be seen in the Figure~\ref{fgr:CI} and relative energies are summarized in the Table~S2. For the triad rsFastLime -- Dronpa -- rsKame the conical intersection energies are inversely proportional to their isomerization quantum yields (and the photoswitching speed). Namely, rsFastLime is the fastest switcher of all three and respectively it's CI has the lowest energy (60.2~kcal/mol), then follows Dronpa (61.6~kcal/mol) and rsKame -- the slowest switcher -- has the highest CI energy of all three (66.7~kcal/mol). On the basis of the results for these three proteins we could have concluded that the energy of CI guides the photoswitching speed. Following this conclusion we would have expected the CI energy of Padron0.9 and bsDronpa be even lower than rsFastLime energy, since both of them exhibit extremely fast switching behavior. However, following numerous attempts on CI optimization, it was not possible to lower the CI energies of Padron0.9 and bsDronpa and they have stayed at 78.3kcal/mol and 76.3 kcal/mol respectively. Hence, not only kinetics but also topology of the CI region is important for photoswitching.

To characterize the CI region we have also identified a local minimum which would be the closest to the CI point. So, for each minimization the structure identified as a CI point was a starting structure and minimization proceeded with no constraints.
For rsFastLime -- Dronpa -- rsKame an energy minimum was found in a vicinity of a CI point with energies of 54.4, 55.9 and 65.1 kcal/mol respectively. Energy minima for Padron0.9 and bsDronpa were identified at 60.6 and 56.5 kcal/mol respectively. However, after visual inspection and comparison of the ground state energies it was established that the minimized structures belong to Franck-Condon region. In order to quantify the extent of the isomerization reaction coordinate we have used calculated the RMSD of the chromophore from the initial minimum ground state structure. RMSD comparison can be found in the Table~S2 and superimposed chromophore structures from FC and CI regions are shown in the Figure S4.  Compiling all of the findings it can be deduced that the CI region of rsFastLime -- Dronpa --rsKame is separated by an energy barrier from a Franck-Condon region, while in case of Padron0.9 and bsDronpa it can be accessed directly. Our results are in accordance with the potential energy surface described by Polyakov et al.\cite{polyakov_potential_2010} for a HBDI chromophore. Authors indicate a possibility of a higher-lying CI with and a low excited-state transition state. At the same time they identify a low-lying CI and a high energy transition state. As such the fast switching of bsDronpa and Padron0.9 can be explained by a flat potential energy surface which directly connects FC and CI regions.




\section{Conclusion}

By selecting a consistent set of reversibly-switchable fluorescent proteins (Dronpa -- rsFastLime -- rsKame -- Padron -- bsDronpa) we have built a protocol completely reproducing their absorption and emission trend. With only a slight blue-shift of only about 2 kcal/mol it provides a solid tool for rational fluorescent protein design and a mechanistic study.

A thorough investigation of an excited-state reaction path allowed us to rationalize photoswitching properties. Via charge transfer analysis and individual aminoacid contributions it has been shown that absorption is more sensitive to mutations and changes in the protein environment than emission. A general cause of this phenomena we see in higher susceptibility of the excited state species. Sometimes denominated as "hot" species \cite{vengris_contrasting_2004}, which are unequilibrated with the protein environment. During the excitation the system is in equilibrium in the ground and not excited state. While during the emission system founds itself in equilibrium in the excited state. Hence environmental changes disturb the system more during the absorption and less during the emission. bsDronpa with its remarkable Stokes shift serves as an example of the protein environment flexibility. The most representative population was identified during the MD simulations. The major difference from the rest of the proteins is n absence of a hydrogen bond between Arg~89 and the imidazoline ring of the chromophore. Our individual aminoacid calculations clearly show Arg~89 contributing to a blue shift in absorption, which is reproduced in bsDronpa.
Further on, an excited state reaction pathway is assessed for a set of photoswitching proteins. It is shown how conical intersection controls the speed of photoswitching as well as brightness. However, it has to be noted that an existence of an excited state energy barrier also play an important role in the protein's photophysics. In general, a higher lying conical intersection is separated by an excited state energy barrier from Franck-Condon region. However, in case of the fastest switching proteins (Padron and bsDronpa) no such barrier was identified. And although conical intersection in these two proteins has a higher energy than in the slower switching ones, it is the flat potential energy surface which facilitates the isomerization. A more accessible conical intersection relates to a lower brightness and a higher isomerization speed. 

Further characterization of the conical intersection topology, such as it's slopeness, as well as excited state classical trajectory calculations could provide more mechanistic details. 

\section{Computational Methods}

\textit{Molecular Dynamics (MD) and Replica-Exchange MD (REMD)}

As the first step vast conformational sampling of all five proteins has been performed using Gromacs 5.0.1\cite{pronk_gromacs_2013} package and the Amber99sb-ILDN forcefield\cite{case_amber_2014}. In case of Dronpa and Padron0.9 X-ray structures of the bright states were used (2Z10\cite{mizuno_light-dependent_2008} and 3ZUJ\cite{regis_faro_low-temperature_2011} respectively). rsKame, rsFastLime and bsDronpa were modelled by point mutations on the basis of Dronpa crystal structure. The list of mutations differentiating Dronpa from the rest of the proteins is given in the Table \ref{tab:mutations}.

In detail description of MD simulation for Dronpa, rsFastLime and rsKame can be found in our previous work\cite{smyrnova_molecular_2015, smyrnova_thermal_2016}. A similar scheme has been adopted for Padron0.9 and bsDronpa. PROPKA \cite{li_very_2005} predictions were used to determine protonation states of the titratable residues. In case of Padron0.9 we have adopted the protonation scheme proposed by Brakemann et al.\cite{brakemann_molecular_2010}, where GLU 144 is protonated and HIS 193 is neutral with a proton placed on N$\epsilon$. Crystal waters present in the X-ray structure were conserved. Then proteins were solved in a 80.5$\times$80.5$\times$80.5 {\AA}$^3$ TIP3P water box and neutralized by Na\textsuperscript{+} and Cl\textsuperscript{-} counterions added at the concentration of 0.15M. Prepared systems were minimized and equilibrated at 300K temperature and 1bar pressure for 100ps. Long-range interactions were included using Particle Mesh Ewald method, cut-off distance was set 13.5\AA. Since a good quality crystal structure was available for Padron0.9, five 100~ns MD runs. were enough for sufficient conformational sampling. A time step of 2~fs and SHAKE algorithm\cite{ryckaert_numerical_1977} were employed. The temperature was controlled using velocity rescaling with a stochastic term\cite{bussi_canonical_2007}.
bsDronpa model has been constructed by introduction of 6 point-mutations (listed in the Table 1) to Dronpa structure. In order to assure a thorough conformational search a Temperature Replica-Exchange MD (T-REMD) simulation was performed. Twenty structures were assigned temperature values from 300K to 330 K which were obtained using the temperature generator server for REMD-simulations\cite{patriksson_temperature_2008}. Equilibration was performed at each of 20 temperatures following the same procedure as described for MD simulation. Then 50~ns runs were performed for each replica thus leading to 1$\mu$s cumulative production run. The resulting exchange probability was $\approx$ 20$\%$ with exchange attempts occurring every 2~ps.

GROMACS utilities were used to perform cluster analysis for each of five proteins. As a results the median structure (the most representative for each simulation) was determined. Then these structures were used as input structures for QM/MM simulations.
\newline
\textit{The QM/MM calculations} 

MOLCAS/Tinker interface\cite{aquilante_molcas_2010} was used to perform QM/MM calculations. MOLCAS 7.8\cite{aquilante_molcas_2010} was used to describe QM part and Tinker 5.1\cite{ponder_efficient_1987} for MM part. In preparation of the QM/MM models the Automated Rhodopsin Modeling (ARM)\cite{melaccio_towards_2016} approach was used. This tool has proved to be a robust protocol for simulations of rhodopsin-like photoreceptors. Certainly there is structural difference between Dronpa-like fluorescent proteins and rhodopsin-like photoreceptors. However similarity in their photobiological characteristics allows to employ the same approach.

The QM/MM model was constructed on the basis of a median structure obtained from classical MD simulations. Proteins were re-centered in the water box, then a sphere of 30{\AA} radius from the chromophore was defined. Since no periodic boundary conditions are employed during the QM/MM simulations this approach allows to save computing time without losing the environment effect. QM/MM model can be divided into three parts: environment, cavity and QM part. Schematic representation of the QM/MM models constructed for five proteins is shown in the Figure~S5.

The input structure used for QM/MM model was subject to 10 $\times$1~ns MD, where the environment was kept frozen and cavity atoms flexible. Then 10 random frames were picked up from the resulting trajectories and used for consecutive QM/MM calculations. Thus for each protein 10 independent QM/MM models were prepared.

\begin{figure}[H]
\includegraphics[width=13.5cm]{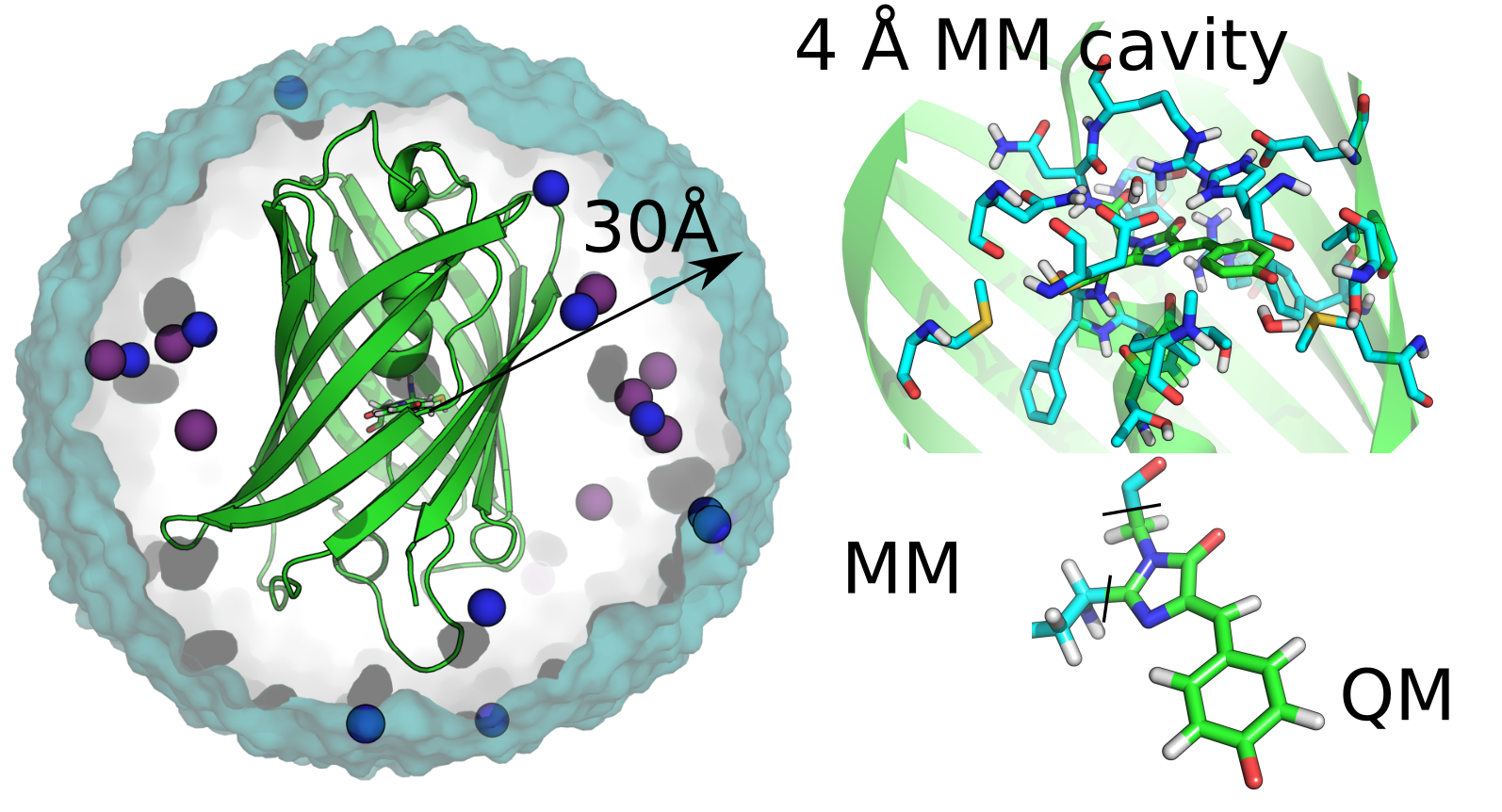}
  \caption{Full system is shown on the left (protein shown as cartoon, Na+  and Cl- ions as spheres (blue and violet respectively), dark cyan surface indicates a 30{\AA} water sphere and chromophore is shown as sticks). A 4{\AA} radius cavity around the chromophore was defined and comprised of X residues (cyan) including the chromophore (green). All this residues were flexible during the optimization. The water molecules present in the crystal structure were maintained and included in the cavity. Chromophore was included in the QM region.}
  \label{fgr:7}
\end{figure}

All the atoms corresponding to the environment are fixed. The 4{\AA} radius cavity was defined around and including the chromophore. Residues in the cavity were free to relax. The QM part comprised of the chromophore's $\pi$-system with link atoms\cite{singh_combined_1986} placed between carbon atoms (shown in the Fig. X). Link atom position was restrained according to Morokuma scheme\cite{svensson_oniom:_1996}. The charge of the frontier atoms was set to 0 and the residual fractional charge redistributed of the rest of chromophore atoms according to atomic mass and insuring -1 integer charge of the chromophore. In all calculations MM atoms were described with Amber94\cite{cornell_second_1995} force field.

As the first step structures were optimized at HF/3-21G/MM level. Then a single-point CASSCF(12,11)/6-31G*/MM calculation was performed in order to define orbitals, which correspond to $\pi$-orbitals and excluding the ones with population close to 0 or 2. Orbitals defined in single-point calculation were used as input for the consecutive CASSCF(12,11)/6-31G*/MM optimization. Then we switch to a CASPT2(12,11)/6-31G*/MM 3-single root single point calculation is performed, taking a 3-roots State Average (SA)\cite{ferre_approximate_2002} CASSCF(12,11)/6-31G*/MM wavefunction as reference. As a result 10 vertical excitation energies ($\Delta$E\textsubscript{S1-S0}) were computed. Then CASSCF(12,11)/6-31G*/MM excited-state optimization was performed. For the located excited-state minimum structure a single-point CASPT2(12,11)/6-31G*/MM 3-single root calculation was performed. Emission energies were computed as the difference between the first two roots ($\Delta$E$'$\textsubscript{S1-S0}). Average over 10 values for each protein was taken as a reference for comparison to the experimental values. 

To calculate the standard deviation from the experimental values we have used the standard formula given in Eq.\ref{eqn:STD} 
\begin{equation}
  \sigma = \sqrt{\frac{1}{N} \sum_{i=1}^N (x_i - x_{exp})^2}\label{eqn:STD}
\end{equation}
where N=10 and x\textsubscript{exp} is an experimental value.

Charge re-distribution shown in the Figure \ref{fgr:charge_redistribution}  was calculated as a difference between the Mulliken charges in the ground and excited states. Hydrogen charges were summed up to the heavy atoms. 

To calculate the effect of each aminoacid on absorption and emission maxima the most representative structures for each protein were used for single-point CASPT2(12,11)/6-31G*/MM 3-single root calculation. Charges of atoms corresponding to each amino acid were set 0 in Tinker key-file.

\begin{acknowledgement}
Financial support from the Flemish Government through the KU Leuven concerted action scheme is gratefully acknowledged.
The computational resources and services used in this work were provided by the VSC (Flemish  Supercomputer Center), funded by the Hercules Foundation and the Flemish Government.

\end{acknowledgement}

\begin{suppinfo}
Following information can be found in the Supporting Information: 
computational details,.

\end{suppinfo}
\bibliography{Better_bibtex}

\end{document}